\documentclass{anmatuvt}
\usepackage{mathrsfs, amssymb, eucal, amsmath, amsthm, amscd, amsrefs, graphicx}
\numberwithin{equation}{section}
\newtheorem{thm}{\bf Theorem}[section]

\newtheorem{defn}{\bf Definition}[section]
\theoremstyle{remark}

\begin{document}
\title[Logical Modelling of Physarum Polycephalum]{Logical Modelling of Physarum Polycephalum}

\author[Schumann]{Andrew Schumann}
\address{Department of Philosophy and Science Methodology \\
Belarusian State University \\
Kalvarijskaya street nr. 9 \\
Minsk \\
Belarus} \email{Andrew.Schumann@gmail.com}

\author[Adamatzky]{Andrew Adamatzky}
\address{Department of Computer Science \\
UWE \\
Bristol \\
United Kingdom} \email{andrew.adamatzky@uwe.ac.uk}

\date{ }

\maketitle


\begin{abstract}
In the paper we proposed a novel model of unconventional computing
where a structural part of computation is presented by dynamics of
Plasmodium of Physarum polycephalum, a large single cell. We
sketch a new logical approach combining conventional logic with
process calculus to demonstrate how to employ formal methods in
design of unconventional computing media presented by Physarum
polycephalum.
\end{abstract}

\section{Introduction}

In the paper we are demonstrating how to design unconventional
computing media taking into account the problem that in
unconventional computing media both structural parts and computing
data of computers are variable ones, in particular in
reaction-diffusion processors \cite{Adamatzky05} we are dealing
with, both the data and the results of the computation are encoded
as concentration profiles of the reagents (on the contrary, in
conventional models of computation one makes a distinction between
the structural part of a computer, which is fixed, and the data
which are variable and on which the computer operates). As a
result, while in conventional models circuits are fixed, in
unconventional computing media circuits could be set up just as
dynamically variable ones within spatio-temporal logic. Solving
this task allows us to build up nature-inspired computer models
and to consider biological and physical systems as computational
models.

One of the unconventional, nature inspired models similar to
reaction-diffusion computing is chemical machine in that molecules
are viewed as computational processes supplemented with a minimal
reaction kinetics. Berry and Boudol first built up a chemical
abstract machine \cite{Berry92} as an example of how a chemical
paradigm of the interactions between molecules can be utilized in
concurrent computations (in algebraic process calculi). We are
considering another abstract machine of reaction-diffusion
computing exemplified by dynamics of Plasmodium of Physarum
polycephalum. This machine is constructed by using process
algebra, too. Using it, we are studying a possibility of logical
representation of the computation in reaction-diffusion systems.
Using notions of space-time trajectories of local domains of a
reaction-diffusion medium we will could the spatial logic of
trajectories, where well-formed formulas and their truth-values
are defined in the unconventional way

Experimental studies and designs of reaction-diffusion computers
could be traced back to the pioneer discovery of L. Kuhnert
(1986). He demonstrated that some very basic image transformations
can be implemented in the light-sensitive Belousov-Zhabotinsky
system. The ideas by L. Kuhnert., K. L. Agladze, I. Krinsky (1989)
on image and planar shape transformations in two-dimensional
excitable chemical media were further developed and modified by N.
G. Rambidi (1998). At that time, during the mid and late nineties,
a range of chemical logical gates were experimentally built in the
Showalter and Yoshikawa laboratories. The first chemical
laboratory prototypes of precipitating chemical processors for
computational geometry were developed by A. Adamatzky (1996). He
also designed and studied a range of hexagonal cellular-automaton
models of reaction-diffusion excitable chemical systems. Now
reaction-diffusion computing is an extremely wide area of
researches towards unconventional computing.

\begin{figure}
\begin{center}
\fbox{\includegraphics[width=0.68\textwidth]{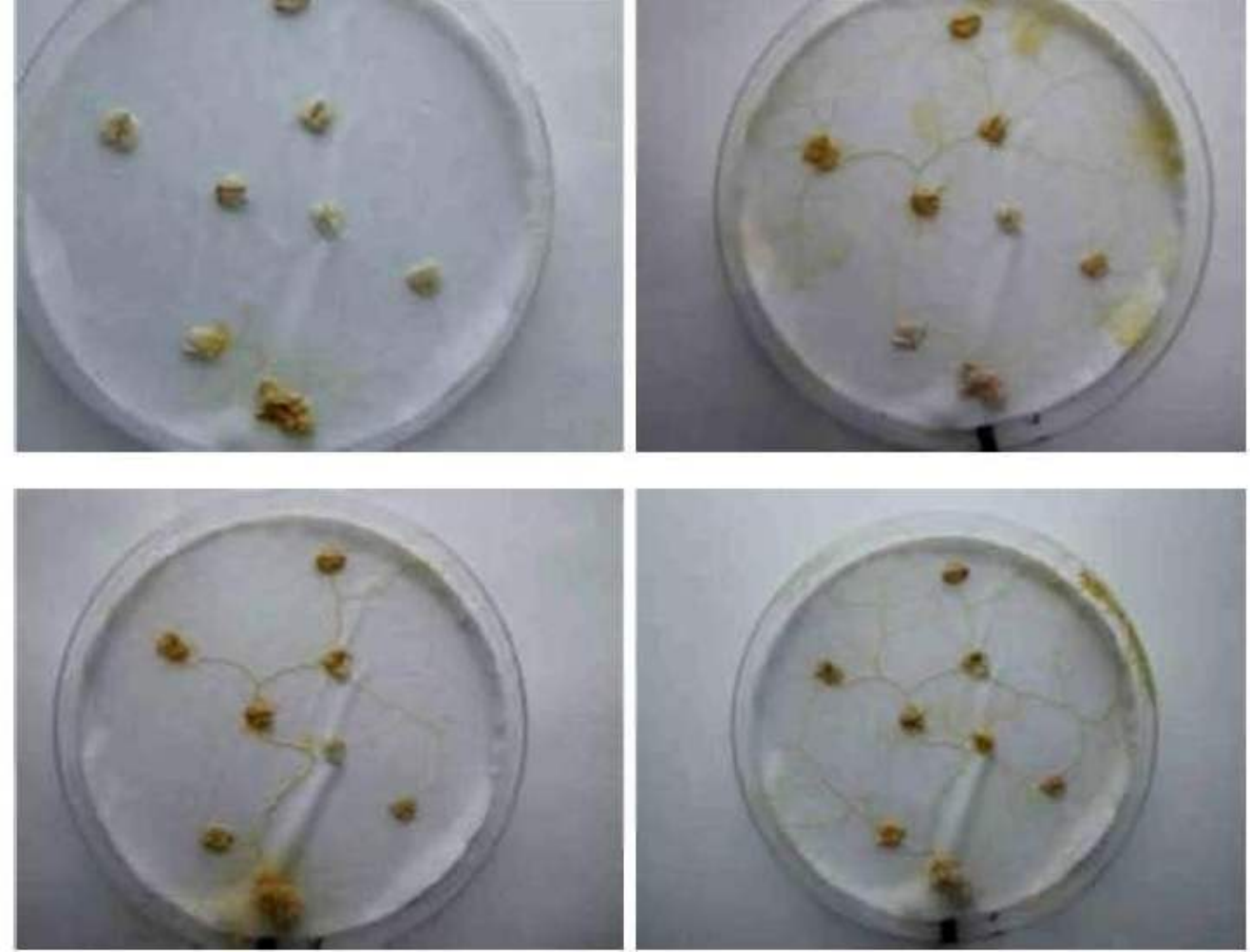}}
\caption{An example of computational process in Physarum machine.
Photographs are taken with time lapse circa 24 hours.}
\label{fig:map1}
\end{center}
\end{figure}

\begin{figure}
\begin{center}
\fbox{\includegraphics[width=0.6\textwidth]{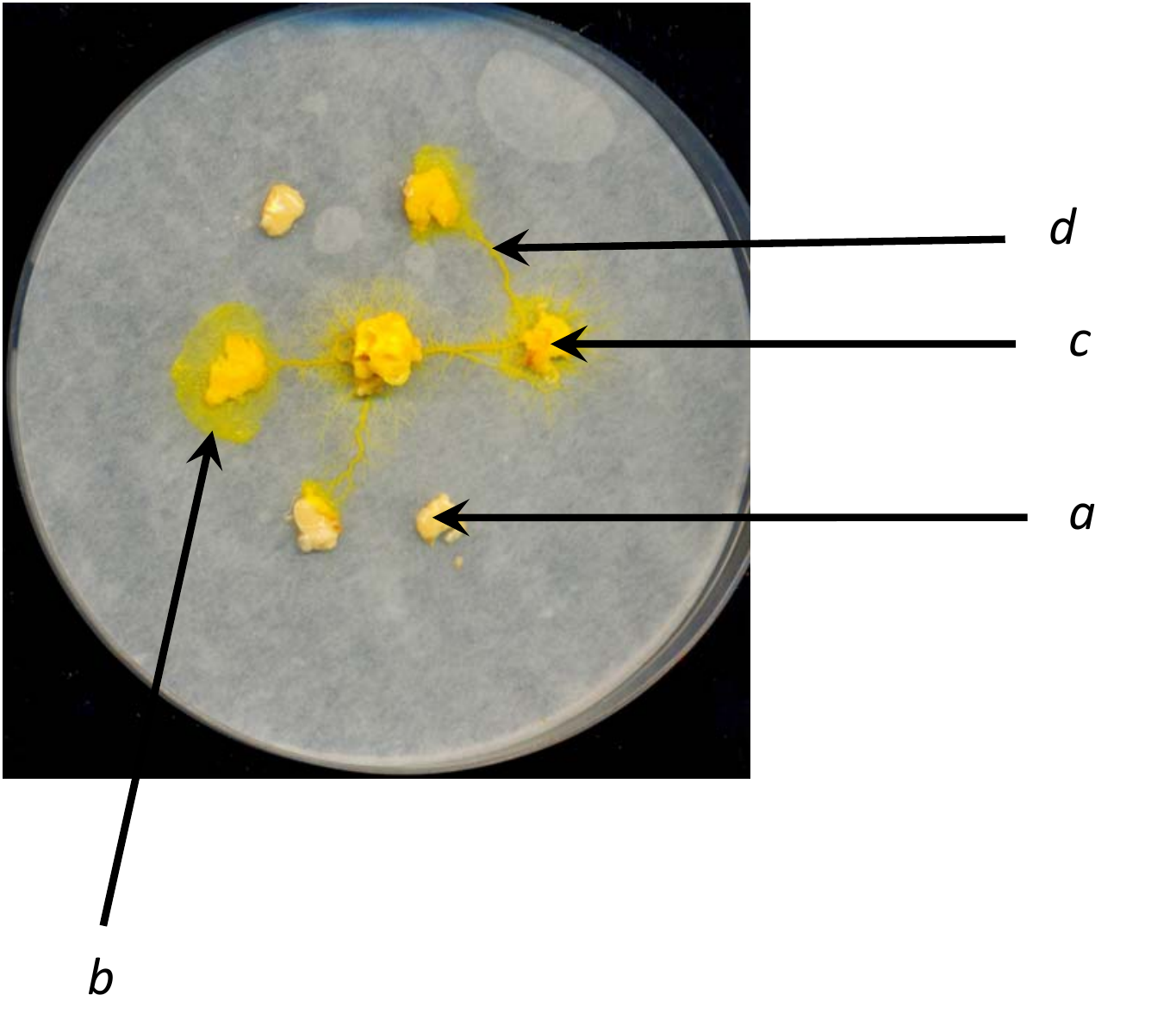}}
\caption{Basic components of Physarum and its environment: (a)~oat
flake, (b)~propagating pseudopodium, plasmodium's wave-fragment,
(c)~oat flake colonized by plasmodium, (d)~protoplasmic tube.}
\label{fig:map2}
\end{center}
\end{figure}

\begin{figure}
\begin{center}
\fbox{\includegraphics[width=0.6\textwidth]{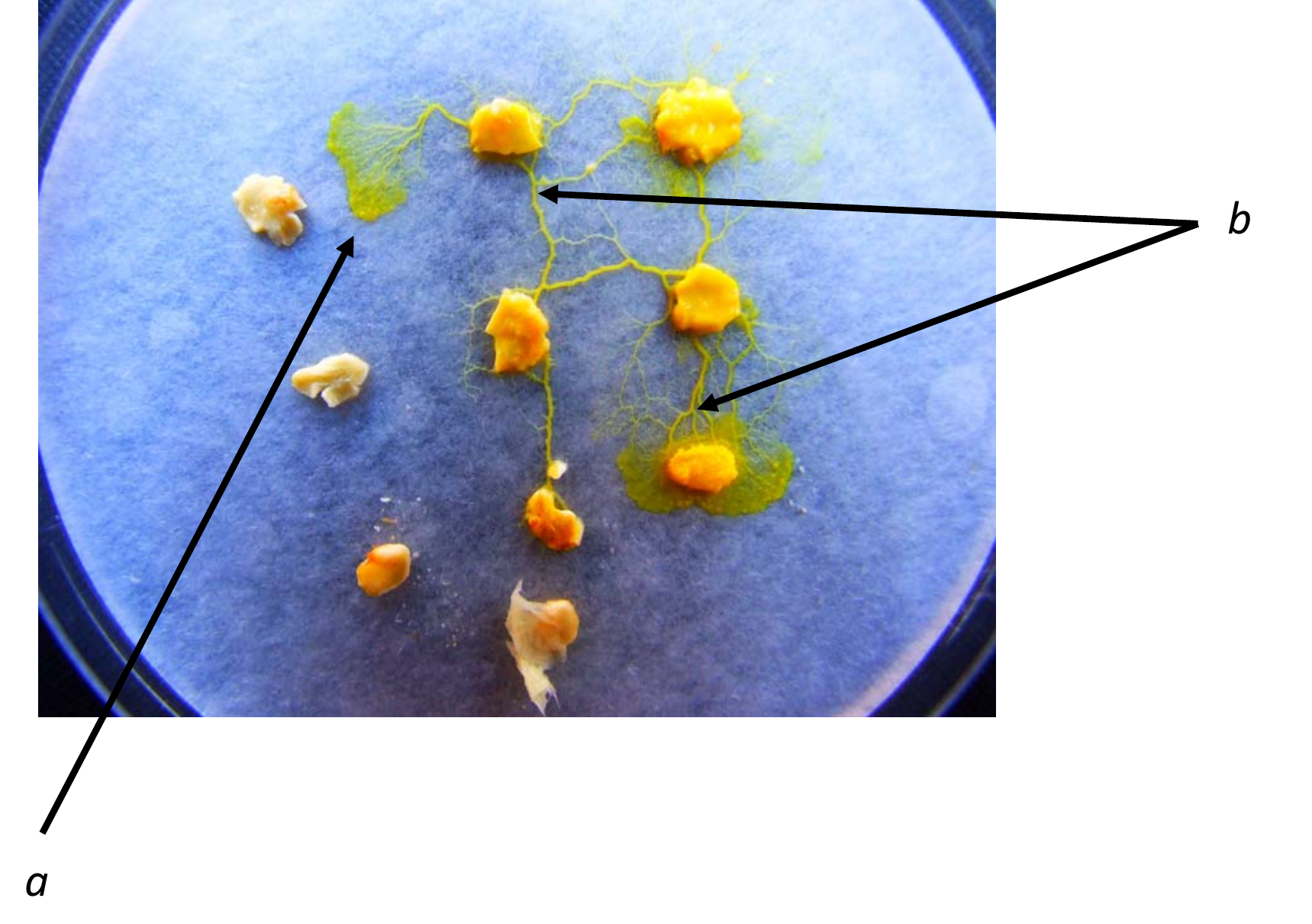}}
\caption{Snapshot of experimental dish with propagating
plasmodium, new activated propagating zone (a) and sites of
branching pseudopodia, junctions of protoplasmic tubes (b) are
shown by arrows.} \label{fig:map3}
\end{center}
\end{figure}

The dynamics of plasmodium of Physarum polycephalum could be
regarded as one of the natural reaction-diffusion computers. The
point is that when the plasmodium is cultivated on a nutrient-rich
substrate (agar gel containing crushed oat flakes) it exhibits
uniform circular growth similar to the excitation waves in the
excitable Belousov-Zhabotinsky medium (Fig.~\ref{fig:map1}). If
the growth substrate lacks nutrients, e.g.\ the plasmodium is
cultivated on a non-nutrient and repellent containing gel, a wet
filter paper or even glass surface localizations emerge and
branching patterns become clearly visible
(Fig.~\ref{fig:map2},~\ref{fig:map3}).

The plasmodium continues its spreading, reconfiguration and
development till there are enough nutrients. When the supply of
nutrients is over, the plasmodium either switches to
fructification state (if level of illumination is high enough),
when sporangia are produced, or forms sclerotium (encapsulates
itself in hard membrane), if in darkness.

The pseudopodium propagates in a manner analogous to the formation
of wave-fragments in sub-excitable Belousov-Zhabotinsky systems.
Starting in the initial conditions the plasmodium exhibits
foraging behavior, searching for sources of nutrients
(Fig.~\ref{fig:map1}). When such sources are located and taken
over, the plasmodium forms characteristic veins of protoplasm,
which contracts periodically. Belousov-Zhabotinsky reaction and
plasmodium are light-sensitive, which gives us the means to
program them. Physarum exhibits articulated negative phototaxis,
Belousov-Zhabotinsky reaction is inhibited by light. Therefore by
using masks of illumination one can control dynamics of
localizations in these media: change a signal's trajectory or even
stop a signal's propagation, amplify the signal, generate trains
of signals. Light-sensitive of Plasmodium has been already
explored in design of robotics controllers \cite{Adamatzky94},
\cite{Adamatzky09}.

Despite numerous experimental implementations of Physarum
computers there is a lack of formalization of the plasmodium's
behavior abstract enough to infer high-level principle of
information transmission by the plasmodium and accurate enough to
reflect peculiarities of the Physarum foraging behavior. In the
paper we are trying to fill the gap and offer interpretation of
Physarum behavior in a framework of process calculi. In the paper
we are trying to define Physarum machine as a process calculus
built on an unconventional interpretation of logical connectives.

\section{A logical approach to analyzing Physarum machine}

Let us set up the problem how to define logical connectives in
Physarum machine. The matter is that both structural parts and
computing data of Physarum computers are variable.

Physarum machine may be viewed as a labelled transition system,
which consists of a collection of states $\mathcal{L} =
\{p_{ij}\}_{i= \overline{1,N},j= \overline{1,K}}$ and a collection
$\mathcal{T}$ of transitions (processes, actions) over them.
Assume $\mathcal{T}\colon \mathcal{L} \mapsto
\mathcal{P}(\mathcal{L})$, where
$\mathcal{P}(\mathcal{L})=\{T\colon T \subseteq \mathcal{L}\}$.
This means that $\mathcal{T}(p)$ consists of all states that a
reachable from $p$. The transition system is understood as a
triple $$\langle \mathcal{L}, \mathcal{T},
\longrightarrow\rangle,$$ where $\longrightarrow \subseteq
\mathcal{L} \times \mathcal{T}\times \mathcal{L}$ is a transition
relation that models how a state $p \in \mathcal{L}$ can evolve
into another state $p'\in \mathcal{L}$ due to an interaction
$\sigma \in \mathcal{T}$. Usually, $\langle p, \sigma, p' \rangle
\in \longrightarrow$ is denoted by $p
\stackrel{\sigma}\longrightarrow p'$. So, a state $p'$ is
reachable from a state $p$ if $p\stackrel{\sigma}\longrightarrow
p'$.

The finite word $\alpha_1 \alpha_2 \dots \alpha_n$ is a
\emph{finite trace of transition system} whenever there is a
finite execution fragment of transition system

$$\varrho = p_0 \alpha_1 p_1 \alpha_2 \dots \alpha_n p_n \,\,\,such\,\, that\,\,\, p_i
\stackrel{\alpha_{i+1}}\longrightarrow p_{i+1}\,\,\, for\,\,
all\,\,\, 0 \leq i < n.$$

The word $\alpha_1 \alpha_2 \dots \alpha_n$ is denoted by
$trace(\varrho)$. The infinite word $\alpha_1 \alpha_2 \dots $ is
an infinite trace whenever there is an infinite execution fragment
of of transition system

$$\varrho = p_0 \alpha_1 p_1 \alpha_2 p_2\alpha_3 p_3\dots \,\,\,such\,\, that\,\,\, p_i
\stackrel{\alpha_{i+1}}\longrightarrow p_{i+1}\,\,\, for\,\,
all\,\,\, 0 \leq i.$$

The word $\alpha_1 \alpha_2 \dots $ is denoted by $trace(\varrho)$
too.
\begin{defn} An infinite (resp. finite) trace of state $p$ denoted by $\varrho(p)$ is the trace of an
infinite (resp. finite) execution fragment starting in $p$.
\end{defn}

Each trace can be regarded as a graph, where nodes represent
states and edges transitions. In this way, transition system is
viewed as graph trees.

Conventionally, logical connectives are defined in the algebraic
way that is broken within transition systems. Therefore in
Physarum machine we will distinguish two kinds of logical
connectives:
\begin{enumerate}
    \item \emph{logical connectives defined co-algebraically},
    these ones are closed to conventional logical connectives:
    they are ``eternal,'' because they are defined over traces (i.e.\ for any future states);
    \item \emph{logical connectives defined as transitions over states}, these ones differ from conventional logical connectives:
    they are defined over states, therefore their values could change for some future states.
\end{enumerate}

\subsection{Logical connectives defined co-algebraically}

First, let us consider logical connectives defined
co-algebraically, i.e.\ by coinduction.

An infinite trace of state $\varrho(p)$ may be presented as a kind
of stream. For a trace $\varrho(p)$, we call $\varrho_p(0)$ the
initial value of $\varrho(p)$. We define the \emph{derivative}
$\varrho_p(0)$ of a trace $\varrho(p)$, for all $n \geq 0$, by
$\varrho_p'(n) = \varrho_p(n + 1)$. For any $n \geq 0$,
$\varrho_p(n)$ is called the $n$-th state of $\varrho(p)$. It can
also be expressed in terms of higher-order trajectory derivatives,
defined, for all $k \geq 0$, by $\varrho_p^{(0)} = \varrho(p)$;
$\varrho_p^{(k+1)} = (\varrho_p^{(k)})'$. In this case the $n$-th
state of a trace $\varrho(p)$ is given by $\varrho_p(n) =
\varrho_p^{(n)}(0)$. So, the trajectory is understood as an
infinite sequence of derivatives: $\varrho(p) = \varrho_p(0) ::
\varrho_p(1) :: \varrho_p(2) :: \dots :: \varrho_p(n-1) ::
\varrho_p^{(n)},$ or $\varrho(p) = \langle\varrho_p(0),
\varrho_p(1), \varrho_p(2), \dots\rangle.$

A \emph{bisimulation} on the set of traces is a relation $R$ such
that, for all $\varrho(p)$ and $\varrho(q)$, if
$\langle\varrho(p), \varrho(q)\rangle\in R$ then (i) $\varrho_p(0)
= \varrho_q (0)$ (this means that they have the same \emph{initial
value}) and (ii) $\langle\varrho_p ', \varrho_q'\rangle \in R$
(this means that they have the same \emph{differential equation}).

If there \emph{exists} a bisimulation relation $R$ with
$\langle\varrho_p, \varrho_q\rangle \in R$ then we write
$\varrho_p \sim\varrho_q$ and say that $\varrho_p$ and $\varrho_q$
are \emph{bisimilar}. In other words, the \emph{bisimilarity}
relation $\sim$ is the greatest bisimulation. In addition, the
bisimilarity relation is an equivalence relation.

\begin{thm}[Coinduction] For all $\varrho(p), \varrho(q)$, if
there exists a bisimulation relation $R$  with $\langle\varrho(p),
\varrho(q)\rangle \in R$, then $\varrho(p) = \varrho(q)$.
\hfill$\Box$
\end{thm}

This proof principle is called \emph{coinduction}. It is a
systematic way of proving the statement using bisimularity:
instead of proving only the single identity $\varrho(p) =
\varrho(q)$, one computes the greatest bisimulation relation $R$
that contains the pair $\langle\varrho(p), \varrho(q)\rangle$. By
coinduction, it follows that $\varrho(p) = \varrho(q)$ for all
pairs $\langle\varrho(p), \varrho(q)\rangle\in R$.

Now consider logical connectives defined by coinduction over
traces. Their syntax is as follows:
\\

\emph{Variables}: $\mathbf{p} ::= p \,\,\,|\,\,\, q \,\,\,| \,\,\,
r \,\,\, \dots$,
\\

\noindent where $p$, $q$, $r$ are states of Physarum machine
presented as a labelled transition system.
\\

\emph{Constants}: $\mathbf{c} ::= \top\,\,\,|\,\,\,\bot$
\\

\noindent where $\top$ means the truth (the ideal, universal
trace) and $\bot$ means the falsity (the empty, impossible trace).
\\

\emph{Formulas}: $\varphi, \psi ::= \textbf{p} \,\,\,|\,\,\,
\mathbf{c} \,\,\,|\,\,\, \neg \psi \,\,\,|\,\,\, \varphi\vee \psi
\,\,\,|\,\,\, \varphi\wedge\psi\,\,\,|\,\,\,\varphi\supset\psi$
\\

These definitions are coinductive. For instance,

\begin{itemize}
    \item a variable $\mathbf{p}$ is of the form of a trace $ \mathbf{p}= \mathbf{p}(0) :: \mathbf{p}(1) ::
\mathbf{p}(2) :: \dots :: \mathbf{p}(n-1) :: \mathbf{p}^{(n)}$,
where $\mathbf{p}(i)\in \{p,q,r,\dots\}$ for each $i\in \omega$;
    \item a constant $\mathbf{c}$
is of the form of a trace $\mathbf{c} = \mathbf{c}(0) ::
\mathbf{c}(1) :: \mathbf{c}(2) :: \dots :: \mathbf{c}(n-1) ::
\mathbf{c}^{(n)}$, where $\mathbf{c}(i)\in \{\top, \bot\}$ for
each $i\in \omega$, a particular case is $[\top] = [\top](0) ::
[\top](1) :: [\top](2) :: \dots :: [\top]^{(n)}$, where
$[\top](i)=\top$ for each $i\in \omega$;
    \item a formula $\neg\varphi$ has the differential
equation $ (\neg\varphi)' = \neg(\varphi')$ and its initial value
is $(\neg\varphi)(0) = \neg\varphi(0)$, this formula will be
understood as $\varphi \supset [\bot]$;
    \item a formula $\varphi\vee\psi$ has the differential
equation $ (\varphi\vee\psi)' = \varphi'\vee\psi'$ and its initial
value is $(\varphi\vee\psi)(0) = \varphi(0)\vee\psi(0)$;
    \item a formula $\varphi\wedge\psi$ has the differential
equation $ (\varphi\wedge\psi)' = \varphi'\wedge\psi'$ and its
initial value is $(\varphi\wedge\psi)(0) =
\varphi(0)\wedge\psi(0)$;
    \item a formula $\varphi\supset\psi$ has the differential
equation $ (\varphi\supset\psi)' = \varphi'\supset\psi'$ and its
initial value is $(\varphi\supset\psi)(0) =
\varphi(0)\supset\psi(0)$.
\end{itemize}

\subsection{Logical connectives defined as transitions over states}

In analyzing the plasmodium we observe processes of inaction,
fusion, and choice, which could be interpreted as unconventional
(spatial) falsity, conjunction and disjunction respectively,
denoted by $Nil$, $\&$ and $+$. These operations differ from
conventional ones, because they cannot have a denotational
semantics in the standard way. However, they may be described as
special transitions over states of Physarum machine:

\begin{enumerate}
    \item inaction ($Nil$) means that pseudopodia has just stopped to behave,
    \item fusion ($\&$) means that two pseudopodia come in contact one with another and then
    merge,
    \item choice (+) means a competition between
two pseudopodia in their bahaviours.
\end{enumerate}

Let us notice that a Boolean algebra may be extended up to the
case of the system of logical connectives defined by coinduction
\cite{Schumann08} (see the previous subsection). However, if we
define three basic logical connectives (falsity, conjunction,
disjunction) as transitions over states of Physarum machine, they
will be extremely non-classical and Boolean properties do not hold
for them in general case.

\section{Physarum process calculus}

Further, let us try to build up a process calculus combining two
approaches to logical connectives for describing the dynamics of
Physarum machine, i.e.\ we are trying to show that indeed this
machine could be presented as a labelled transition system with
some logical relations.

Assume that the computational domain $\Omega$  is partitioned into
computational cells $c_j = 1, \dots, K$ such that $c_i \cap c_j =
\emptyset$, $i \neq j$ and $\bigcup_{j=1}^{K}c_j =\Omega$. Then
suppose that in the $K$ cells, there are $N$ active species or
growing pseudopodia and the state of species $i$ in cell $j$ is
denoted by $p_{ij}$, $i = 1, \dots, N$, $j = 1, \dots, K$. These
states are time dependent and they are changed by plasmodium's
active zones interacting with each other and affected by
attractants or repellents. Plasmodium's active zones interact
concurrently and in a parallel manner. Foraging plasmodium can be
represented as a set of following abstract entities
(Fig.~\ref{fig:map2}).

\begin{enumerate}
    \item The set of \emph{active zones} (growing pseudopodia or actions) $Z =
\{a, b, \dots\}$ (Fig. 2a). On a nutrient-rich substrate
plasmodium propagates as a typical circular, target, wave, while
on the nutrient-poor substrates localized wave-fragments are
formed. Each action $a$ from $Z$ belongs to a state $p_{ij}$, $i =
1, \dots, N$, $j = 1, \dots, K$ of a cell $i$, which is its
current position, and says about a transition (propagation) of a
state $p_{ij}$ to another state of the same or another cell. Part
of plasmodium feeding on a source of nutrients may not propagate,
so its transition is nil, but this part can always start moving.
    \item The set of \emph{attractants} $\{A_1, A_2, \dots\}$ are sources of
nutrients, on which the plasmodium feeds. It is still subject of
discussion how exactly plasmodium feels presence of attracts,
indeed diffusion of some kind is involved. Based on our previous
experiments we can assume that if the whole experimental area is
about $8-10\sim cm$ in diameter then the plasmodium can locate and
colonize nearby sources of nutrients. Each attract $A(a)$ is a
function from $a$ to another action $b$.
    \item The set of \emph{repellents} $\{R_1, R_2, \dots\}$. Plasmodium
of Physarum avoids light. Thus, domains of high illumination are
repellents such that each repellent R is characterized by its
position and intensity of illumination, or force of repelling. In
other words, each repellent $R(a)$ is a function from $a$ to
another action $b$.
    \item The set of \emph{protoplasmic tubes} $\{C_1, C_2, \dots\}$.
Typically plasmodium spans sources of nutrients with protoplasmic
tubes/veins (Fig.~\ref{fig:map2}). The plasmodium builds a planar
graph, where nodes are sources of nutrients, e.g.\ oat flakes, and
edges are protoplasmic tubes. $C(a)$ means a diffusion of growing
pseudopodia by an action $a$.
\end{enumerate}

Our process calculus contains the following basic operators:
`$Nil$' (inaction), `$\star$' (prefix), `$|$' (cooperation),
`$\backslash$' (hiding), `$\&$' (reaction/fusion), `$+$' (choice),
$a$ (constant or restriction to a stable state), $A(\cdot)$
(attraction), $R(\cdot)$ (repelling), $C(\cdot)$
(spreading/diffusion). Let $\Lambda = \{a, b, \dots\}$ be a set of
names. With every $a \in\Lambda$ we associate a complementary
action $\overline{a}$. Define $L = \{a, \overline{a} : a\in
\Lambda\}$, where $a$ is considered as \emph{activator} and as
\emph{inhibitor} for $a$, be the set of labels built on $\Lambda$
(under this interpretation, $a = \overline{\overline{a}}$).
Suppose that an action $a$ communicates with its complement
$\overline{a}$ to produce the internal action $\tau$. Define
$L_\tau  = L  \cup \{\tau\}$.

We use the symbols $\alpha$, $\beta$, etc., to range over labels
(actions), with $a = \overline{\overline{a}}$, and the symbols
$P$, $Q$, etc., to range over processes on states $p_{ij}$, $i =
1, \dots, N$, $j = 1, \dots, K$. The processes are given by the
syntax:
\\

$ P, Q ::= Nil ~~~|~~~ \alpha \star P ~~~|~~~ A(\alpha)\star P~~~
|~~~ R(\alpha) \star P ~~~| ~~~C(\alpha)~~~ | ~~~(P|Q)~~~ | ~~~P
\backslash Q ~~~ |$ $~~~~~~~~~~~~P \& Q~~~ | ~~~P + Q ~~~| ~~~a$
\\

Each label is a process, but not vice versa. An operational
semantics for this syntax is defined as follows:

$$ \text{\textbf{Prefix}:}~~~~  \frac{}{\alpha \star P \stackrel{\alpha }\longrightarrow P},$$
$$ \frac{}{A(\alpha) \star P \stackrel{\beta}\longrightarrow P}~~(A(\alpha)=\beta),$$
$$ \frac{}{R(\alpha) \star P \stackrel{\beta}\longrightarrow P}~~(R(\alpha)=\beta),$$

(the conclusion states that the process of the form $\alpha \star
P$ (resp.\ $A(\alpha) \star P$ or $R(\alpha) \star P$) may engage
in $\alpha$ (resp.\ $A(\alpha)$ or $R(\alpha)$) and thereafter
they behave like $P$; in the presentations of behaviors as trees,
$\alpha \star P$ (resp.\ $A(\alpha)\star P$ or $R(\alpha)\star P$)
is understood as an edge with two nodes: $\alpha$ (resp.\
$A(\alpha)$ or $R(\alpha)$) and the first action of $P$),

$$\text{\textbf{Diffusion}:} ~~~\frac{P\stackrel{\alpha }\longrightarrow P'}{P\stackrel{\alpha }\longrightarrow C(\alpha)}~~~(C(\alpha) ::= P'),$$
$$\text{\textbf{Constant}:} ~~~\frac{P\stackrel{\alpha }\longrightarrow P'}{a\stackrel{\alpha }\longrightarrow P'}~~~(a ::= P, a \in L_\tau),$$
$$\text{\textbf{Choice}:}
~~~\frac{P\stackrel{\alpha }\longrightarrow
P'}{P+Q\stackrel{\alpha }\longrightarrow
P'},~~~\frac{Q\stackrel{\alpha }\longrightarrow
Q'}{P+Q\stackrel{\alpha }\longrightarrow Q'},$$

(these both rules state that a system of the form $P + Q$ saves
the transitions of its subsystems $P$ and $Q$),

$$\text{\textbf{Cooperation}:} ~~~\frac{P\stackrel{\alpha }\longrightarrow P'}{P| Q\stackrel{\alpha }\longrightarrow
P'|Q},~~~\frac{Q\stackrel{\alpha }\longrightarrow
Q'}{P|Q\stackrel{\alpha }\longrightarrow P|Q'},$$

(according to these rules, the cooperation | interleaves the
transitions of its subsystems),

$$\frac{P\stackrel{\alpha }\longrightarrow
P'\qquad Q\stackrel{\overline{\alpha}}\longrightarrow Q'
}{P|Q\stackrel{\tau}\longrightarrow P'|Q'},$$

(i.e. subsystems may synchronize in the internal action $\tau$ on
complementary actions $\alpha$ and $\alpha$),

$$\text{\textbf{Hiding}:} ~~~\frac{P\stackrel{\alpha }\longrightarrow P'}{P\backslash Q\stackrel{\alpha }\longrightarrow
P'\backslash Q}~~(\alpha \notin Q, Q \subseteq  L),$$

(this rule allows actions not mentioned in $Q$ to be performed by
$P \backslash Q$),

$$\text{\textbf{Fusion}:}~~~\frac{}{\alpha \star P\& \overline{P} \stackrel{\alpha }\longrightarrow Nil}$$

(the fusion of complementary processes are to be performed into
the inaction),

\[\frac{P\stackrel{\alpha }{\longrightarrow} P'\begin{array}{cc} {} & {} \end{array} Q\stackrel{\alpha }{\longrightarrow} P'}{P\& Q\stackrel{\alpha }{\longrightarrow} P'} ,\qquad \frac{P\stackrel{\alpha }{\longrightarrow} P'\begin{array}{cc} {} & {} \end{array}Q\stackrel{\alpha }{\longrightarrow} P'}{Q\& P\stackrel{\alpha }{\longrightarrow} P'} \]
(this means that if we obtain the same result $P'$ that is
produced by the same action $\alpha$ and evaluates from two
different processes $P$ and $Q$, then $P'$ may be obtained by that
action $\alpha$ started from the fusion $P \& Q$ or $Q \& P$),

\[\frac{P\stackrel{\alpha }{\longrightarrow} P'}{P\& Q\stackrel{\alpha }{\longrightarrow} Nil+ C(\alpha )+ P'} , \qquad\frac{ P\stackrel{\alpha }{\longrightarrow} P'}{Q\& P\stackrel{\alpha }{\longrightarrow} Nil+ C(\alpha )+ P'} \]
(these rules state that if the result $P'$ is produced by the
action $\alpha$ from the processes $P$, then a fusion $P \& Q$ (or
$Q \& P$) is transformed by that same $\alpha$ either into the
inaction or  diffusion or process $P'$).

These are inference rules for basic operations. The ternary
relation $P \stackrel{\alpha }{\longrightarrow}  P'$ means that
the initial action $P$ is capable of engaging in action $\alpha$
and then behaving like $P'$.

The informal meanings of basic operations are as follows:

\begin{enumerate}
    \item $Nil$, this is the empty process which does nothing. In other
words, $Nil$ represents the component which is not capable of
performing any activities: a deadlocked component.
    \item $\alpha \star P$, a process $\alpha \in L$ followed by the
process $P$: $P$ becomes active only after the action $\alpha$ has
been performed. An activator $\alpha \in L$ followed by the
process $P$ is interpreted as branching pseudopodia into two or
more pseudopodia, when the site of branching represents newly
formed process $\alpha \star P$.

In turn, an inhibitor $\overline{\alpha }\in L$ followed by the
process $P$ is annihilating protoplasmic strands forming a process
at their intersection.
    \item $A(\alpha)\star P$ denotes a process that
waits for a value $\alpha$ and then continues as $P$. This means
that an attractor $A$ modifies propagation vector of action
$\alpha$ towards $P$. Attractants are sources of nutrients. When
such a source is colonized by plasmodium the nutrients are
exhausted and attracts ceases to function: $A(\alpha)\star Nil$.
    \item $R(\alpha)\star P$ denotes a process that
waits for a value $\alpha$ and then continues as $P$. This means
that a repellent $R$ modifies propagation vector of action
$\alpha$ towards $P$. Process can be cancelled, or annihilated, by
a repellent: $R(\alpha)\star Nil$. This happens when propagating
localized pseudopodium $\alpha$ enters the domain of repellent,
e.g.\ illuminated domain, and $\alpha$ does not have a chance to
divert or split.

    \item $C(\alpha)$, a diffusion of activator
$\alpha\in L$ is observed in placing sources of nutrients nearby
the protoplasmic tubes belonging to $\alpha$ or inactive zone
($\alpha ::= Nil$). More precisely, diffusion generates
propagating processes which establish a protoplasm vein (the case
of activator $\alpha$) or annihilate it (when source of nutrients
exhausted, the case of inhibitor $\overline{\alpha }$).

    \item $P |  Q$, this is a parallel composition
(commutative and associative) of actions: $P$ and $Q$ are
performed in parallel. The parallel composition may appear in the
case, two more food sources are added to either side of the array
and then the plasmodium sends two streams outwards to engulf the
sources. When the food sources have been engulfed, the plasmodium
shifts in position by redistributing its component parts to cover
the area created by the addition of the two new processes $P$ and
$Q$ that will already behave in parallel.

Process $P$ can be split, or multiplied, by two sources of
attractants $(A_1A_2)(P) \star P_1 |  P_2$. Pseudopodium $P$
approaches the site where distance to $A_1$ is the same as
distance to $A_2$. Then $P$ subdivides itself onto two pseudopodia
$P_1$ and $P_2$. Each of the pseudopodia travels to its unique
source of attractants. Also, process $P$ can be split, or
multiplied, by a repellent: $R(P) \star P_1 |  P_2$. Biophysics of
fission with illuminated geometrical shapes is discussed in [5].
The fission happens when a propagating pseudopodium `hits' a
repellent. The part of pseudopodium most affected by the repellent
ceases propagating, while two distant parts continue their
development. Thus, two separate pseudopodia are formed.

    \item $P \backslash  Q$, this restriction operator allows
us to force some of $P$'s actions not to occur; all of the actions
in the set $Q \subseteq L$ are prohibited, i.e.\ the component $P
\backslash Q$ behaves as P except that any activities of types
within the set $Q$ are hidden, meaning that their type is not
visible outside the component upon completion.

    \item $P \& Q$, this is the fusion of $P$ and $Q$; $P \& Q$
represents a system which may behave as both component $P$ and
$Q$. For instance, $Nil$ behaves as $P \& \overline{P}$, where $P$
is an activator and $\overline{P}$ an appropriate inhibitor
respectively. The fusion of $P$ and $Q$ is understood as collision
of two active zones $P$ and $Q$. When they collide they fuse and
annihilate, $P \& Q \star Nil$. Depending on the particular
circumstances the new active zone $\alpha$ (the result of fusing)
may become inactive ($Nil$), transform to protoplasmic tubes
($C(\alpha)$), or remain active and continue propagation in a new
direction (the case of prefix $\star$).

When two pseudopodia come in contact one with another, they do
usually merge (Fig.~\ref{merging}). Thus by directing processes
with attractors we can merge the processes: $A(P_1, P_2) \star P_1
\& P_2$ (see details in [5]).

\begin{figure}
\begin{center}
\fbox{\includegraphics[width=0.4\textwidth]{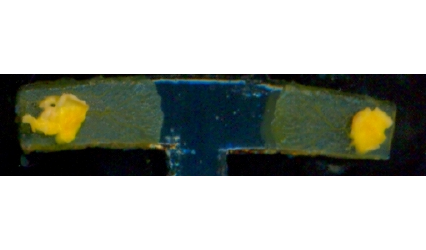}}\qquad\fbox{\includegraphics[width=0.37\textwidth]{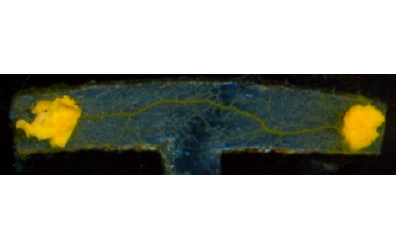}}
\caption{Merging of two plasmodium's wave-fronts. Photos are made
with interval 9 hours.} \label{merging}
\end{center}
\end{figure}

    \item $P + Q$, this is the choice between $P$ and $Q$; $P + Q$
represents a system which may behave either as component $P$ or as
$Q$. Thus the first activity to complete identifies one of the
components which is selected as the component that continues to
evolve; the other component is discarded. In Physarum calculi, the
choice $P + Q$ between processes $P$ and $Q$ sometimes is
represented by competition between pseudopodia tubes $C(P)$ and
$C(Q)$, i.e.\ $C(P, Q) = C(P) + C(Q)$. In other words, two
processes $P$ and $Q$ can compete with each, during this
competition one process `pulls' protoplasm from another process,
thus making this another process inactive. The competition happens
via protoplasmic tube.

    \item $a$, constants belonging to labels
are components whose meaning is given by equations such as $a ::=
P$. Here the constant $a$ is given the behaviour of the component
$P$. Constants can be used to describe infinite behaviours, via
mutually recursive defining equations.

\end{enumerate}

Thus, in this process calculus we have two kinds of logical
connectives.

\begin{enumerate}
    \item The group of connectives defined by coinduction. They are
derivable from the hiding. Indeed, let 1 be a universal set of
active zones, then the following equalities hold:
$$\neg P ::= 1 \backslash  P  ~~~~~ \text{\textbf{negation}},$$
$$P\wedge Q ::= P \backslash  (1
\backslash  Q)  ~~~~~ \text{\textbf{conjunction}},$$
$$P \vee Q ::= 1 \backslash  ((1
\backslash P)\backslash  Q)  ~~~~~ \text{\textbf{disjunction}},$$
$$P \supset Q ::= 1 \backslash  (P
\backslash  Q)  ~~~~~ \text{\textbf{implication}}.$$ These
connectives satisfy all properties of Boolean algebra.
    \item The group of connectives defined as transitions. It
    consists of three operations: \textbf{inaction}, \textbf{fusion} and \textbf{choice}.
    Their basic properties:
\begin{equation}Nil \backslash  P \cong Nil,\end{equation}
\begin{equation}P \& \overline{P}\cong Nil,\end{equation}
\begin{equation}P \& P \cong P,\end{equation}
\begin{equation}P \& Nil \cong Nil,\end{equation}
\begin{equation}(P + Q) \backslash  P' \cong P \backslash  P' + Q
\backslash  P',\end{equation}
\begin{equation} (P \& Q) \backslash  P' \cong P \backslash  P' \& Q \backslash  P',\end{equation}
\begin{equation} P \& Q \cong Q \& P,\end{equation}
\begin{equation} P \& (Q \& R) \cong (P \& Q) \& R,\end{equation}
\begin{equation} P + P \cong P,\end{equation}
\begin{equation} P + Nil \cong P,\end{equation}
\begin{equation} P + Q \cong Q + P,\end{equation}
\begin{equation} P + (Q + R) \cong (P + Q) + R,\end{equation}
\begin{equation} P \& (Q + R) \cong (P \& Q) + (P \& R),\end{equation}
\begin{equation} P + (Q \& R) \cong (P + Q) \& (P + R),\end{equation}
where $\cong$ is a congruence relation defined on the set of
processes.
\end{enumerate}

\section{ Conclusion}

In the paper we have just shown that the behavior of plasmodium of
Physarum polycephalum could be considered as a kind of process
calculus with several logical connectives defined in non-standard
way. Thus, the media of Physarum polycephalum can be viewed as one
of the natural unconventional (reaction-diffusion) computers. Its
weakest point is that \emph{the speed of computation is so slow}:
each new state of Physarum dynamics may be observed just in hours
(see Fig.~\ref{fig:map1}).


\begin{bibdiv}
\begin{biblist}
\bib{Adamatzky05}{book}{
title={Reaction-Diffusion Computers}, author={Adamatzky, A.},
author={De Lacy Costello, B.}, author={Asai, T.}, date={2005},
publisher={Elsevier}, address={Amsterdam} }

\bib{Adamatzky94}{article}{title={Physarum machines: encapsulating reaction-diffusion
to compute spanning tree}, author={Adamatzky, Andrew},
journal={Naturwisseschaften}, volume={94}, date={2007},
pages={975--980}}

\bib{Adamatzky09}{article}{title={Steering Plasmodium with light: Dynamical programming of Physarum machine}, author={Adamatzky, Andrew}, journal={New Mathematics and Natural Computation}, date={2009}}

\bib{Berry92}{article}{title={The chemical abstract machine}, author={Berry, G.}, author={Boudol, G.}, journal={Theor. Comput.
Sci.}, volume={96}, date={1992}, pages={217–-248}}

\bib{Schumann08}{article}{title={Non-well-founded Probabilities and Coinductive
Probability Logic}, author={Schumann, Andrew}, journal={Eighth
International Symposium on Symbolic and Numeric Algorithms for
Scientific Computing (SYNASC'08)}, date={2008}, pages={54--57} }

\bib{Schumann09}{article}{title={Towards Semantical Model of
Reaction-Diffusion Computing}, author={Schumann, Andrew},
author={Adamatzky, Andrew}, journal={Kybernetes}, volume={38/9},
date={2009}, pages={1518--1531} }
\end{biblist}
\end{bibdiv}
\end{document}